**Magnetoelectrically driven catalytic degradation of organics**

*Fajer Mushtaq\*, Xiang-Zhong Chen\*, Harun Torlakcik, Christian Steuer, Marcus Hoop, Erdem Can Siringil, Xavi Marti, Gregory Limburg, Patrick Stipp, Bradley J. Nelson and Salvador Pané\**

F. Mushtaq, Dr. X.-Z. Chen, H. Torlakcik, Dr. M. Hoop, E. C. Siringil, G. Limburg, P. Stipp, Prof. Dr. B. J. Nelson, Dr. S. Pané

Multi-Scale Robotics Lab (MSRL), Institute of Robotics and Intelligent Systems (IRIS), ETH Zurich, CH-8092 Zurich, Switzerland.

E-mail: fmushtaq@ethz.ch, chenxian@ethz.ch, vidalp@ethz.ch

Dr. C. Steuer
Institute of Pharmaceutical Sciences, ETH Zurich, CH-8093 Zurich, Switzerland.

Dr. Xavi Marti
Institute of Physics, Academy of Sciences of the Czech Republic, Cukrovarnick´a 10, 162 00 Praha 6, Czech Republic
IGS Research Ltd., Calle La Coma, Nave 8, 43140 La Pobla de Mafumet, Tarragona, Spain

**Abstract**

Here, we report the catalytic degradation of organic compounds by exploiting the magnetoelectric (ME) nature of cobalt ferrite-bismuth ferrite (CFO-BFO) core-shell nanoparticles. The combination of magnetostrictive CFO with the multiferroic BFO gives rise to a magnetoelectric engine that purifies water under wireless magnetic fields via advanced oxidation processes, without involvement of any sacrificial molecules or co-catalysts. Magnetostrictive $CoFe_2O_4$ nanoparticles are fabricated using hydrothermal synthesis, followed by sol-gel synthesis to create the multiferroic $BiFeO_3$ shell. We perform theoretical modeling to study the magnetic field induced polarization on the surface of magnetoelectric nanoparticles. The results obtained from these simulations are consistent with the experimental findings of the piezo-force microscopy analysis, where we observe changes in the piezoresponse of the nanoparticles under magnetic fields. Next, we investigate the magnetoelectric effect induced catalytic degradation of organic pollutants under AC magnetic fields and obtained 97% removal efficiency for synthetic dyes and over 85% removal efficiency for routinely used pharmaceuticals. Additionally, we perform trapping experiments to elucidate the mechanism behind the magnetic field induced catalytic degradation of organic pollutants by using scavengers for each of the reactive species. Our results indicate that hydroxyl and superoxide radicals are the main reactive species in the magnetoelectrically induced catalytic degradation of organic compounds.

Magnetic nanostructures have been widely used as magnetically recoverable catalysts or as carriers for catalytic materials.[1] While magnetic nanomaterials have found widespread applications in tuning catalytic processes, a majority of the employed strategies focus on their motion to enhance the reagents' mass transport.[2] A next level of control can be achieved by forcing such magnetic nanoparticles to interact[3] or, alternatively, by coupling them to catalysts to enhance the reaction performance.[4] In previous investigations, magnetic fields only support the catalytic chemical conversion but never act as the ultimate trigger. Direct control of causality is fundamental in realistic scenarios, where the choice of the precise moment of actuation is critical. In this direction, magnetically induced heating has been demonstrated to initiate chemical catalysis on demand.[5] Here, we demonstrate a localized trigger for catalytic reactions via the direct magnetoelectric (ME) effect on the surface of multiferroic nanoparticles. Our ME nanocatalysis is able to decompose organic contaminants, such as dyes and various pharmaceuticals, without the involvement of any sacrificial molecules or co-catalysts. Our experiments revealed a rich interdependence between the applied magnetic field parameters and the reaction speed, which significantly improves the performance of catalytic reactions for environmental remediation.

Organic pollutants such as pharmaceuticals, pesticides and industrial chemicals are persistent compounds that are resistant to degradation through conventional processes and bio-accumulate in ecosystems, causing severe impacts on human health and the environment.[6] Towards this effect, billions of dollars have been invested annually to study new approaches capable of efficiently combating this global crisis.[7] Surface charges that can initiate redox reactions and form hydroxyl and superoxide radicals, can serve as an attractive approach for non-selective degradation of such organic pollutants. The occurrence of surface charges, thus, becomes the trigger to ignite such reactions and shifts the spotlight to any mechanism that can generate such surface charges on demand. The magnetoelectric (ME) effect not only fulfills this requirement, but, also allows for wireless operation while restricting the targeted regions to areas where the ME materials are deployed. While theoretically possible, experimentally realizing this has been hindered by the dearth of ME materials and nanostructures that actually can be fabricated. Many research efforts have responded to this call by investigating the growth, characterization and operation of ME nanoparticles (NPs) in-depth.

Our main results are summarized in **Figure 1**. We have investigated the capability of our core-shell ME NPs to initiate electrochemical processes under the application of alternating magnetic fields (Figure 1a) by studying the degradation of a model organic pollutant, rhodamine B (RhB). Degradation curves obtained for RhB under alternating magnetic fields for CFO-BFO and controls are presented in Figure 1b. From this figure we can observe that the control sample (without any NPs) and the CFO NPs sample displayed a negligible response under alternating magnetic fields. BFO NPs, showed a slight decrease in RhB concentration (22%), which can be attributed to the weak coupling between the ferromagnetic and ferroelectricity of BFO at room temperature.[8] In contrast, core-shell CFO-BFO NPs demonstrated an elevated RhB degradation efficiency of 97% within 50 min (Figure 1b). We were also successful in extending our novel approach for degradation of a cocktail of five commonly used pharmaceuticals, Carbamazepine, Diclofenac, Gabapentin, Oxazepame and Fluconazole[6b, 9]. Figures 1c and S1 show that all five pharmaceuticals were degraded with over 80% efficiency. These results confirm our hypothesis that our ME NPs can be used to simultaneously target a wide variety of organic compounds in a non-selective approach. This further highlights the advantage of using our approach over conventional ozone treatments for pharmaceutical removal in wastewater treatment plants, which is known to display a negligible reactivity towards a vast variety of ozone-resistant such as Gabapentin, Oxazepame and Fluconazole.[10]

The magnetoelectric nanocatalysts used in this work consist of magnetostrictive cobalt (II) ferrite ($CoFe_2O_4$, CFO) cores coated with multiferroic bismuth (III) ferrite ($BiFeO_3$, BFO) shells. CFO NPs were fabricated by a hydrothermal synthesis approach by carefully tuning the growth conditions to obtain single-crystalline and phase-pure NPs. A co-precipitation method was employed to form CFO NPs using NaOH solution to form precipitates of iron and cobalt hydroxides.[11] CTAB was used as the surfactant to control the nucleation and shape of the CFO NPs.[12] This mixture was then sealed in an autoclave and placed in an oven at elevated temperature for hydrothermal treatment. Core-shell magnetoelectric (ME) CFO-BFO NPs were fabricated by coating the CFO NPs with a BFO precursor via a sol-gel approach, followed by annealing the NPs to crystallize the BFO shell.[13] (Figure S2). CFO NPs fabricated in this study are octahedrons and have an average size of 30 nm. (Figure S3 and S4a). The core-shell CFO-BFO NPs have an average size of 42 nm (**Figure 2**a, S4b and S5). Energy-dispersive X-ray (EDX) mappings confirm the presence of a shell composed of bismuth, iron and oxygen around the cobalt ferrite core (Figure 2b, S6 and S7).

The crystalline structure of the CFO and CFO-BFO NPs was analyzed using transmission electron microscopy (TEM) and X-ray diffraction (XRD). XRD patterns (Figure 2c) showed that for CFO NPs all peaks can be assigned to the pure Fd3m structure of $CoFe_2O_4$ (JCPDS No. 01-1121), indicating a cubic spinel structure. Similar analysis performed on CFO-BFO NP sample shows that, in addition to the $CoFe_2O_4$ peaks, they possess new peaks that can be assigned to the pure phase of $BiFeO_3$ (JCPDS No. 71-2494), indicating a rhombohedral perovskite structure with the space group *R3c*. A high resolution TEM (HRTEM) analysis performed on a single CFO NP sample is presented in Figure 2d, featuring an intact and orderly structure. The planes with interplanar d-spacing of 0.295 nm matches the (220) crystal face of CFO. Its corresponding selected area electron diffraction (SAED) pattern is presented in Figure 2e, which indicates the occurrence of a single-crystalline CFO structure. The spots in the SAED pattern have been indexed according to the pure cubic spinel (Fd3m) structure of CoFe$_2$O$_4$. The HRTEM image obtained for the BFO shell shows the presence of an intact, orderly, single-crystalline structure (Figure 2f). The planes with interplanar d-spacing of 0.198 nm match the (024) crystal face. Its corresponding SAED pattern is presented in Figure 2g and shows that the BFO shell is also single crystalline. The spots in the SAED pattern have been indexed according to the *R3c* structure of BiFeO$_3$.

The ferroelectricity and magnetoelectricity of a single core-shell NP was directly probed using piezoresponse force microscopy (PFM) under an external magnetic field. A conductive cantilever tip was used in contact mode to apply an alternating voltage to the CFO-BFO NP and induce piezoelectric surface oscillations, which were sensed through the cantilever deflection. To investigate the ferroelectric and magnetoelectric coupling effect in our CFO-BFO NPs, local piezoresponse hysteresis loops were obtained at random locations of an NP (**Figure 3**a, b) by sweeping the applied DC bias, while simultaneously measuring the phase and amplitude response. The excitation voltage waveform was programmed to be a stepwise increasing pulsed DC voltage that was superimposed on a small AC voltage. In order to minimize the possible interference caused by electrostatic forces, the AC response signal was acquired only during the off-phase of the voltage pulse sequence.[14] From the phase loop presented in Figure 3a it can be clearly observed that the BFO shell exhibits polarization reversibility both with and without the application of the external magnetic field. From this image it is clear that BFO's polarization directions can be switched at both polarities of the tip DC-bias voltage. Both piezoresponse phase loops are horizontally shifted, a trend that can also be observed from the amplitude curves with asymmetric butterfly shape (Figure 3b)[15] This asymmetry in the loops can be attributed to many factors, such as the imprint effect, internal bias fields inside the materials, and/or due to a work function difference between the top, Pt-coated Si probe and the bottom gold electrode.[16] The coercive voltages for the BFO shell measured without magnetic field are -4.69 V and 2.65 V, respectively. When the magnetic field was applied, the coercive voltages changed to -3.47 V and 3.06 V, respectively.

This smaller coercive voltage obtained under a magnetic field indicates that the strain generated in the magnetostrictive CFO core was effectively transferred to the shell, facilitating the polarization reversal process in BFO. This is indirect evidence of strain mediated magnetoelectric effect in the core-shell CFO-BFO NPs. The positive coercive voltage change (0.41 V) is smaller than the negative coercive voltage change (1.22 V). The asymmetric change indicates that there is an offset of the center of the piezoresponse loop under the magnetic field, which is caused by an electric field generated by the magnetoelectric effect.[14b, 17] The magnetoelectric coupling coefficient is defined as,

$$\alpha = \frac{\Delta E}{\Delta H}$$

where $\Delta H$ is the change in magnetic field and $\Delta E$ is the change in the electric field caused by the external magnetic field. For our CFO-BFO NPs, the $\Delta E$ i.e. the offset of center of the loop upon application of magnetic field can be estimated to be (1.22 – 0.41 V)/2/10 nm = 40.5 MV m$^{-1}$. Hence, the local ME coefficient can be estimated as $40.5 \times 10^4$ mV cm$^{-1}$ Oe$^{-1}$. This value is in the same order of magnitude as those reported for some core-shell magnetoelectric nanostructures such as FeGa@P(VDF-TrFE), CoFe$_2$O$_4$-PbZr$_{0.52}$Ti$_{0.48}$O$_3$, CoFe$_2$O$_4$@BaTiO$_3$ and CoFe$_2$O$_4$@BiFeO$_3$, which were evaluated by similar methods.[18] In order to further investigate this magnetoelectric coupling observed in our CFO-BFO NPs, we performed finite element simulations on a single core-shell NP under static magnetic fields. Figures 3c and S7 presents the strain distribution generated on the BFO shell when the core-shell NP was placed in an external magnetic field of 15 mT. From Figures 3d and S8, we can observe the corresponding electric potential gradient induced on the surface of the BFO shell, which is determined by the magnetoelectric coupling and the compliance matrix of the BFO shell. From these simulations we can observe that, when subjected to external magnetic fields, a CFO-BFO NP can generate a local surface potential in the µV range, which can be exploited to initiate certain chemical reactions.

Towards this effect, we used our core-shell NPs to study the degradation trend of the organic pollutant RhB dye under magnetic fields. These degradation curves were already presented in Figure 1b and proved that only the core-shell CFO-BFO NPs were capable of removing 97% of the organic pollutant. This enhanced organic pollutant removal performance of CFO-BFO NPs can be attributed to the ME effect induced redox reactions that are responsible for the catalytic degradation of RhB. Quantitative analysis on ME effect-induced RhB degradation was performed by comparing the reaction rate constants $k$, which can be defined by,

$$k = \ln\frac{C_o}{C} / t \qquad (1)$$

where $C_o$ is the initial RhB concentration and $C$ is the RhB concentration at time $t$. This calculation is based on the assumption that the kinetics of RhB degradation reaction catalyzed by the CFO-BFO nanostructures are (pseudo)-first-order reactions.[19]

The effect of varying magnetic field strengths and frequencies on RhB degradation rate is shown in **Figure 4**a and b. We observed that increasing the magnetic field strength and frequency have a positive effect on RhB degradation rate. These results are also supported by the simulations performed on the core-shell NPs (Figure S10). Based on these results, a magnetic field strength of 15 mT and a frequency of 1.1 kHz were chosen as our preferred magnetic field parameters for further degradation experiments ($k$-value of 0.0725 min$^{-1}$). At these chosen parameters we also investigated the effect of concentration of BFO sol-gel precursor solution on the degradation efficiency of RhB to find the optimal CFO-BFO morphology for further experiments (Figures S11-12).

In order to elucidate the magnetoelectric effect-induced RhB degradation mechanism, we performed trapping experiments of the prominent reactive species that are responsible for decomposition of organic pollutants. For this, degradation of RhB dye was carried out under the optimized magnetic field parameters in the presence of CFO-BFO NPs and different reactive species scavengers (Table 1). It can be seen from **Figure 5**a that the catalytic degradation efficiency decreases with the addition of scavengers, proving that they all participate in the degradation of RhB. Addition of the electron scavenger, $AgNO_3$ and the hole scavenger, ethylene diamine tetraacetic acid (EDTA)[20] lowers the reaction-rate constants. Trapping superoxide radical $O_2^{\cdot-}$ with benzoquinone (BQ) or the hydroxyl radical $OH^{\cdot}$ radical with tert-butyl alcohol (TBA) suppressed the degradation of RhB greatly. These results reveal that the predominant reactive species for magnetoelectrically-induced RhB degradation were the radicals. In addition to the trapping experiments, we confirmed the formation of hydroxyl radicals in our degradation experiments by using terephthalic acid as a photoluminescent $OH^{\cdot}$ trapping agent. Terephthalic acid readily reacts with $OH^{\cdot}$ radicals to produce a fluorescent product, 2-hydroxyterephthalic acid, which emits a fluorescent signal at 425 nm.[20b] From the results of this experiment (Figure 5b) we can observe an increase in fluorescence intensity at 425 nm with increasing piezo-photocatalytic reaction time, which offers further proof of $OH^{\cdot}$ formation during the catalytic reaction.

In this work, we have successfully fabricated core-shell magnetoelectric $CoFe_2O_4$-$BiFeO_3$ NPs and demonstrated their use for catalytic degradation of organic compounds. In a two-step process, magnetostrictive CFO NPs were fabricated using a hydrothermal approach, followed by a sol-gel method to form a BFO shell. The magnetoelectric nature of the NPs has been evidenced by the observed modulation of their piezoelectric response upon the application of magnetic fields. Such findings are consistent with our theoretical modeling of strain mediated magnetoelectric effect in core-shell NPs. Assisted by magnetic fields, our NPs have been able to degrade, first, RhB pollutant with an efficiency of 97% within 50 min and, later, a cocktail comprising routinely employed micro-pollutants in the pharmaceutical industry. To understand the mechanism behind this magnetoelectric effect-induced degradation, trapping experiments have been performed using scavengers for each of the reactive species. Our results indicate that OH$^\bullet$ and O$_2^{\bullet-}$ radicals are the main reactive species in the magnetoelectrically induced catalytic degradation of organic compounds.

**Acknowledgements**

Author Contributions：Conceptualization: F.M., X.C. and S.P.; methodology: F.M., X.C., M.H.; investigation: F.M., X.C., H.T., C.S., E.C.S, G.L., P.S.; writing, original draft: F.M and S.P.; writing, review and editing: F.M., S.P., X.C., X.M., B.N., M.H., H.T., C.S., E.C.S., G.L., and P.S.; supervision: X.C., S.P. and B.N.; funding acquisition: S.P.
This work has been financed by the European Research Council Starting Grant "Magnetoelectric Chemonanorobotics for Chemical and Biomedical Applications (ELECTROCHEMBOTS)", by the ERC grant agreement no. 336456. The authors would like to acknowledge the Scientic Center for Optical and Electron Microscopy (ScopeM) of ETH Zurich, the Institute of Geochemistry and Petrology and the FIRST laboratory, ETH Zurich for their technical support.

**References**

[1] L. M. Rossi, N. J. S. Costa, F. P. Silva, R. Wojcieszak, *Green Chemistry* **2014**, *16*, 2906.

[2] a) K. Ngamchuea, K. Tschulik, R. G. Compton, *Nano Research* **2015**, *8*, 3293; b) C. Niether, S. Faure, A. Bordet, J. Deseure, M. Chatenet, J. Carrey, B. Chaudret, A. Rouet, *Nature Energy* **2018**.

[3] A. Zakharchenko, N. Guz, A. M. Laradji, E. Katz, S. Minko, *Nature Catalysis* **2018**, *1*, 73.

[4] J. Sa, J. Szlachetko, M. Sikora, M. Kavcic, O. V. Safonova, M. Nachtegaal, *Nanoscale* **2013**, *5*, 8462.

[5] a) A. Meffre, B. Mehdaoui, V. Connord, J. Carrey, P. F. Fazzini, S. Lachaize, M. Respaud, B. Chaudret, *Nano Letters* **2015**, *15*, 3241; b) B. Alexis, L. Lise-Marie, F.

Pier-Francesco, C. Julian, S. Katerina, C. Bruno, *Angewandte Chemie International Edition* **2016**, *55*, 15894.

[6] a) L. Rossi, P. Queloz, A. Brovelli, J. Margot, D. A. Barry, *PLOS ONE* **2013**, *8*, e58864; b) L. Kovalova, H. Siegrist, H. Singer, A. Wittmer, C. S. McArdell, *Environmental Science & Technology* **2012**, *46*, 1536.

[7] M. Boehler, B. Zwickenpflug, J. Hollender, T. Ternes, A. Joss, H. Siegrist, *Water Science and Technology* **2012**, *66*, 2115.

[8] a) P. K., S. P. M., S. A., N. T. S., T. T., *physica status solidi (RRL) – Rapid Research Letters* **2012**, *6*, 244; b) C. Ederer, N. A. Spaldin, *Physical Review B* **2005**, *71*, 060401; c) T. E. Quickel, L. T. Schelhas, R. A. Farrell, N. Petkov, V. H. Le, S. H. Tolbert, *Nature Communications* **2015**, *6*, 6562.

[9] L. Kovalova, H. Siegrist, U. von Gunten, J. Eugster, M. Hagenbuch, A. Wittmer, R. Moser, C. S. McArdell, *Environmental Science & Technology* **2013**, *47*, 7899.

[10] Y. Lee, L. Kovalova, C. S. McArdell, U. von Gunten, *Water Research* **2014**, *64*, 134.

[11] a) G. B. Ji, S. L. Tang, S. K. Ren, F. M. Zhang, B. X. Gu, Y. W. Du, *Journal of Crystal Growth* **2004**, *270*, 156; b) E. Pervaiz, I. H. Gul, H. Anwar, *Journal of Superconductivity and Novel Magnetism* **2013**, *26*, 415; c) L. J. Cote, A. S. Teja, A. P. Wilkinson, Z. J. Zhang, *Fluid Phase Equilibria* **2003**, *210*, 307.

[12] a) M. Vadivel, R. R. Babu, K. Ramamurthi, M. Arivanandhan, *Ceramics International* **2016**, *42*, 19320; b) A. L. Lopes-Moriyama, V. Madigou, C. P. d. Souza, C. Leroux, *Powder Technology* **2014**, *256*, 482.

[13] a) Q. Zhang, D. Sando, V. Nagarajan, *Journal of Materials Chemistry C* **2016**, *4*, 4092; b) J. W. Lin, T. Tite, Y. H. Tang, C. S. Lue, Y. M. Chang, J. G. Lin, *Journal of Applied Physics* **2012**, *111*, 07D910; c) K. Chakrabarti, K. Das, B. Sarkar, S. Ghosh, S. K. De, G. Sinha, J. Lahtinen, *Applied Physics Letters* **2012**, *101*, 042401; d) G. F., C. X. Y., Y. K. B., D. S., R. Z. F., Y. F., Y. T., Z. Z. G., L. J.-M., *Advanced Materials* **2007**, *19*, 2889.

[14]  a) X.-Z. Chen, Q. Li, X. Chen, X. Guo, H.-X. Ge, Y. Liu, Q.-D. Shen, *Advanced Functional Materials* **2013**, *23*, 3124; b) X.-Z. Chen, M. Hoop, N. Shamsudhin, T. Huang, B. Özkale, Q. Li, E. Siringil, F. Mushtaq, L. Di Tizio, B. J. Nelson, S. Pané, *Adv. Mater.* **2017**, *29*, 1605458; c) B. F. Mushtaq, X. Chen, M. Hoop, H. Torlakcik, E. Pellicer, J. Sort, C. Gattinoni, B. J. Nelson, S. Pane, *iScience*.

[15]  a) S. H. Xie, J. Y. Li, R. Proksch, Y. M. Liu, Y. C. Zhou, Y. Y. Liu, Y. Ou, L. N. Lan, Y. Qiao, *Applied Physics Letters* **2008**, *93*, 222904; b) G. Caruntu, A. Yourdkhani, M. Vopsaroiu, G. Srinivasan, *Nanoscale* **2012**, *4*, 3218.

[16]  a) X.-Z. Chen, N. Shamsudhin, M. Hoop, R. Pieters, E. Siringil, M. S. Sakar, B. J. Nelson, S. Pane, *Materials Horizons* **2016**, *3*, 113; b) C. Xiang-Zhong, L. Qian, C. Xin, G. Xu, G. Hai-Xiong, L. Yun, S. Qun-Dong, *Advanced Functional Materials* **2013**, *23*, 3124; c) X.-Z. Chen, X. Chen, X. Guo, Y.-S. Cui, Q.-D. Shen, H.-X. Ge, *Nanoscale* **2014**, *6*, 13945.

[17]  H. Miao, X. Zhou, S. Dong, H. Luo, F. Li, *Nanoscale* **2014**, *6*, 8515.

[18]  a) C. Xiang-Zhong, H. Marcus, S. Naveen, H. Tianyun, Ö. Berna, L. Qian, S. Erdem, M. Fajer, D. T. Luca, N. B. J., P. Salvador, *Advanced Materials* **2017**, *29*, 1605458; b) S. Xie, F. Ma, Y. Liu, J. Li, *Nanoscale* **2011**, *3*, 3152; c) F. Bi, L. Ruie, G. Kun, Y. Yaodong, W. Yaping, *EPL (Europhysics Letters)* **2015**, *112*, 27002; d) Q. Zhu, Y. Xie, J. Zhang, Y. Liu, Q. Zhan, H. Miao, S. Xie, *Journal of Materials Research* **2014**, *29*, 657.

[19]  a) F. Mushtaq, A. Asani, M. Hoop, X.-Z. Chen, D. Ahmed, B. J. Nelson, S. Pané, *Adv. Funct. Mater.* **2016**, *26*, 6995; b) F. Mushtaq, M. Guerrero, M. S. Sakar, M. Hoop, A. M. Lindo, J. Sort, X. Chen, B. J. Nelson, E. Pellicer, S. Pané, *J. Mater. Chem. A* **2015**, *3*, 23670.

[20]  a) N. Zhang, D. Chen, F. Niu, S. Wang, L. Qin, Y. Huang, *Scientific Reports* **2016**, *6*, 26467; b) J. Wu, W. Mao, Z. Wu, X. Xu, H. You, A. X. Xue, Y. Jia, *Nanoscale* **2016**, *8*, 7343.

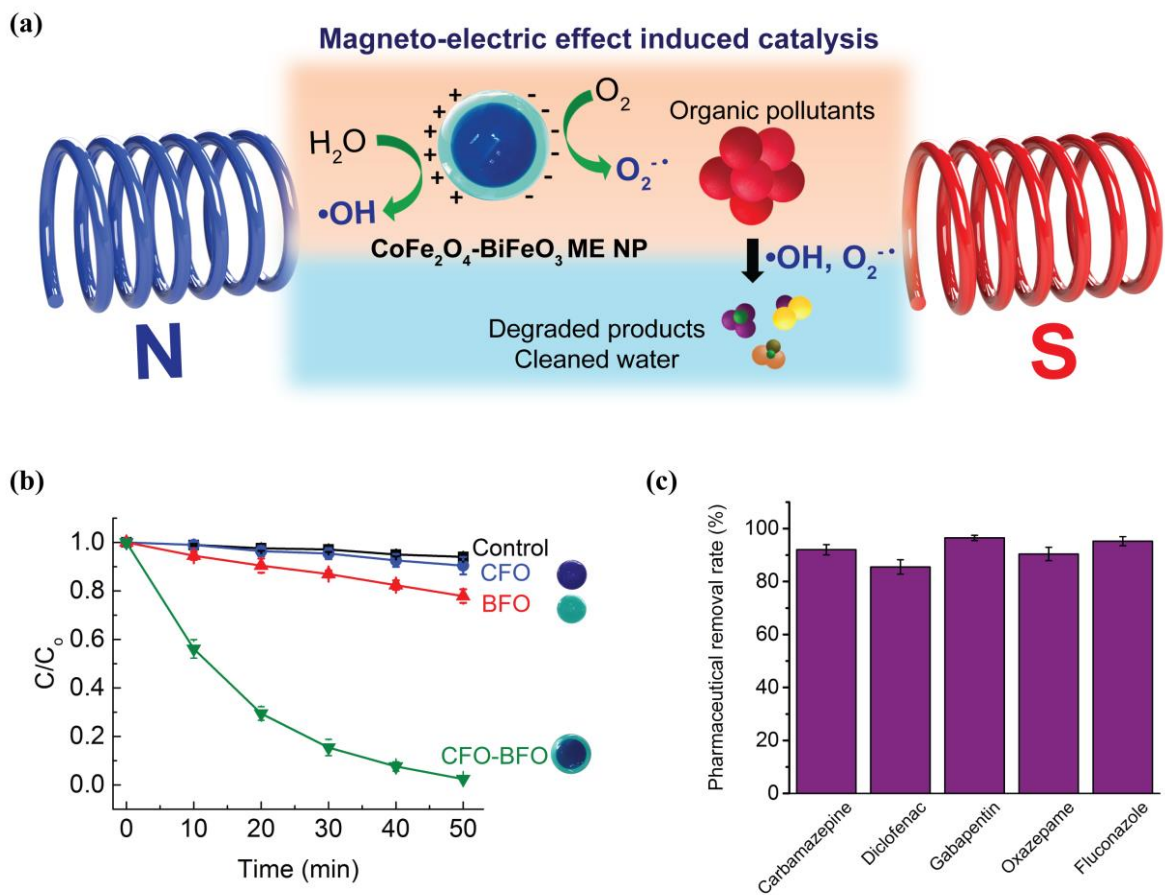

**Figure 1.** (a) Scheme showing magnetoelectric (ME) effect induced catalytic degradation of organic pollutants using core-shell CFO-BFO NPs under magnetic fields. (b) Catalytic degradation curves obtained for model organic dye, RhB, under 15 mT and 1 kHz magnetic fields (n=5). (c) Removal efficiency of a cocktail of five common pharmaceuticals using the core-shell NPs (n=4).

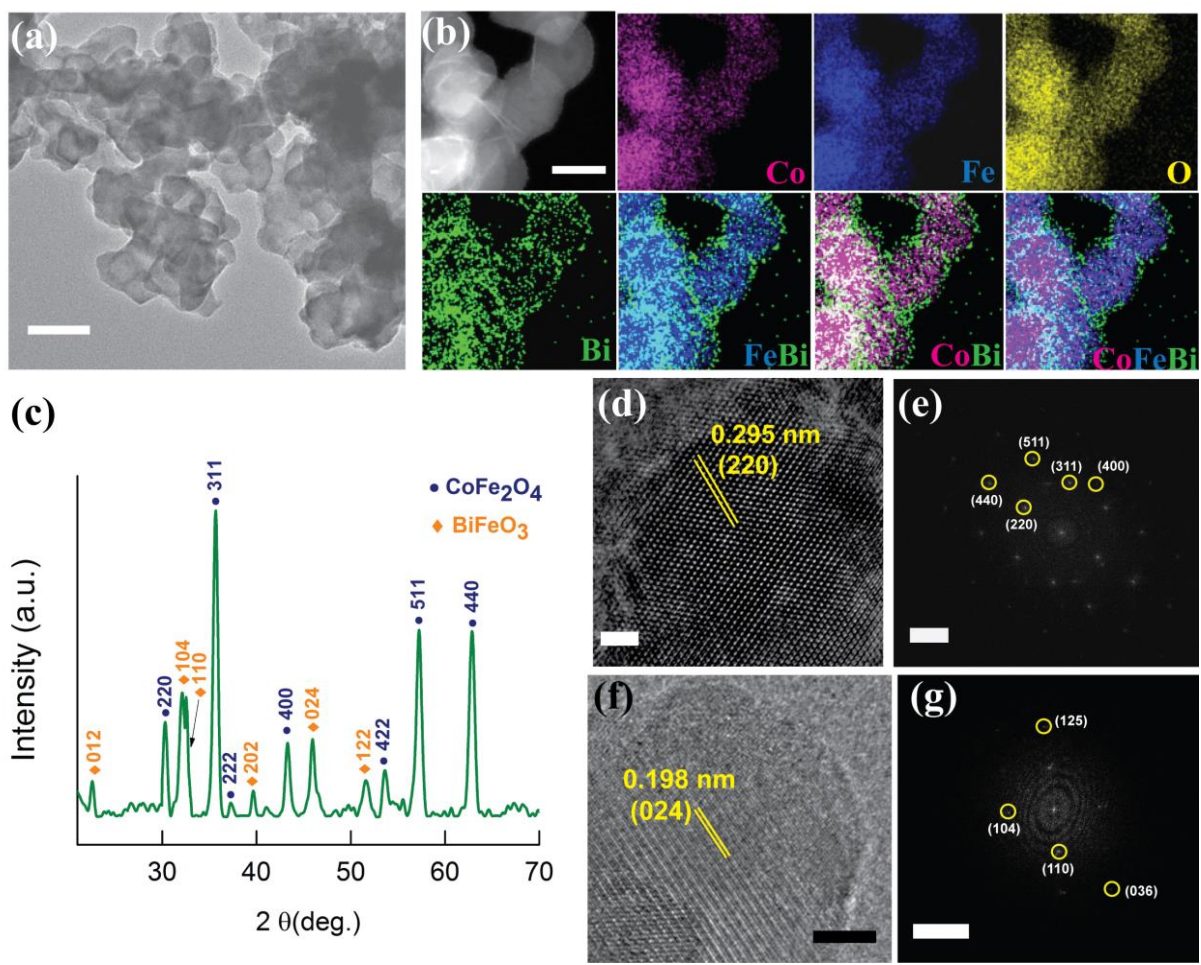

**Figure 2.** Structural characterisation of core-shell CFO-BFO NPs. (a) TEM image showing many overlapped CFO-BFO NPs. (b) HAADF STEM image obtained for some overlapped CFO-BFO NPs and its corresponding EDX maps obtained for Co, Fe, O and Bi, with the superimposed images clearly showing core-shell CFO-BFO NPs. (c) XRD patterns obtained for core-shell NPs. (b) HRTEM image of a single CFO NP (c) and its corresponding SAED pattern. (d) HRTEM image of a core-shell NP showing the BFO shell region and (e) its corresponding SAED pattern. (scale bars: (a) 100 nm, (b) 30 nm, (d) 2 nm, (e) 4 nm$^{-1}$, (f) 2.5 nm and (g) 3 nm$^{-1}$).

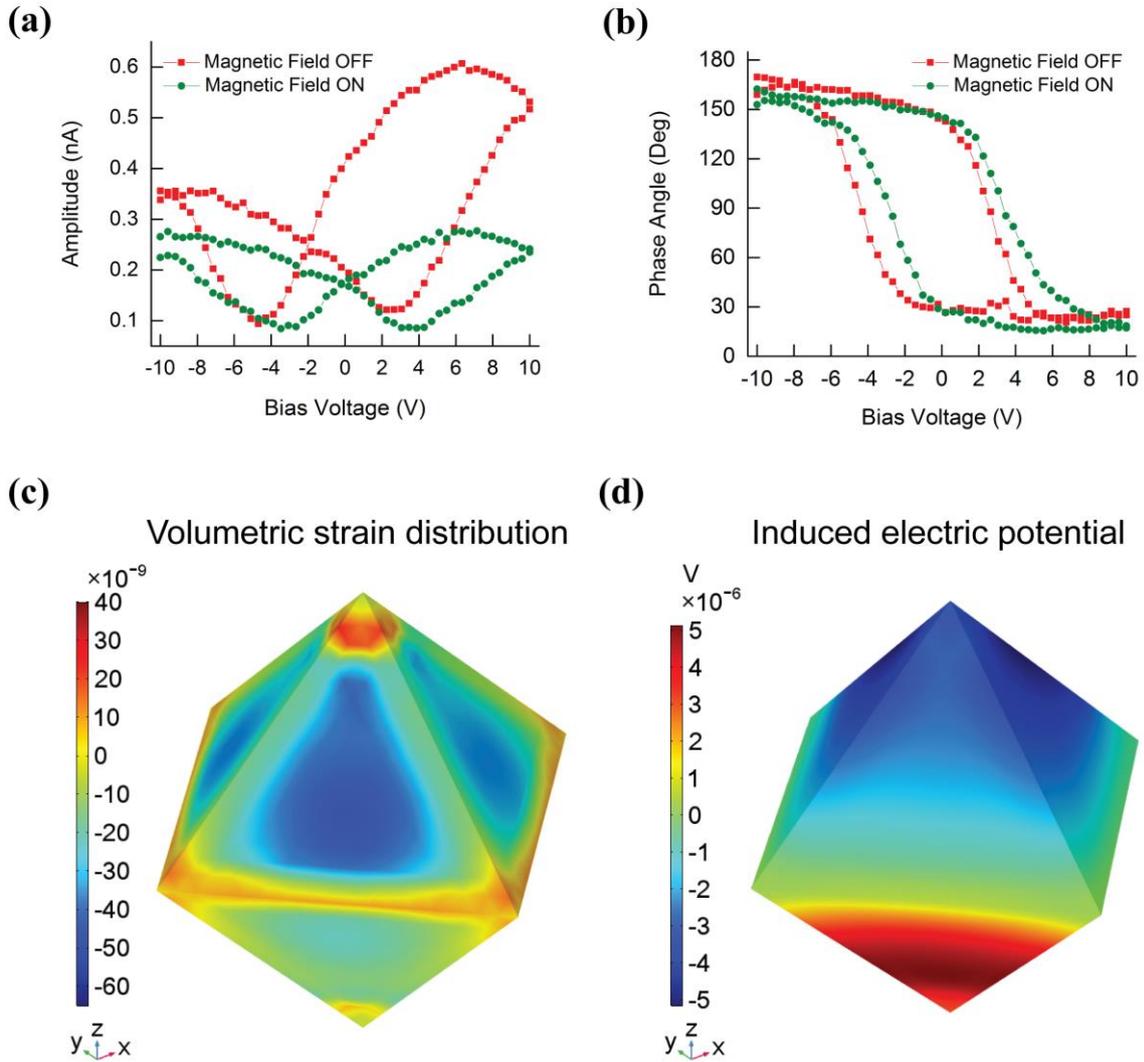

**Figure 3.** Ferroelectric and magnetoelectric characterisation of core-shell CFO-BFO nanostructures. (a) Amplitude response of a single core-shell NP obtained with and without magnetic field and (b) the corresponding phase response. (c) COMSOL simulations performed on a CFO-BFO nanopartcile under a magnetic field of 15 mT showing the (c) strain generated on the BFO shell due to the magnetostrictive CFO core and (d) the corresponding electric potential induced on the surface of the BFO shell.

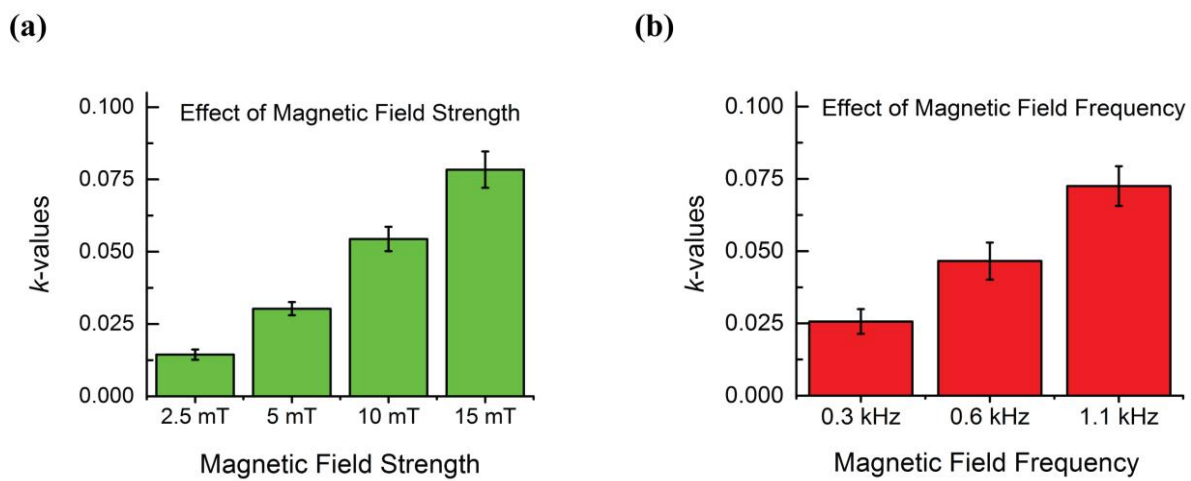

**Figure 4.** Magnetoelectric effect induced catalytic degradation of organic pollutants using core-shell CFO-BFO NPs under alternating magnetic fields. (a) Comparison of RhB degradation rate contants obtained by using CFO-BFO NPs under different magnetic field strengths at 1.1 kHz field frequency (n=5) and (b) under different field frequencies and at a magnetic field strength of 15 mT (n=5).

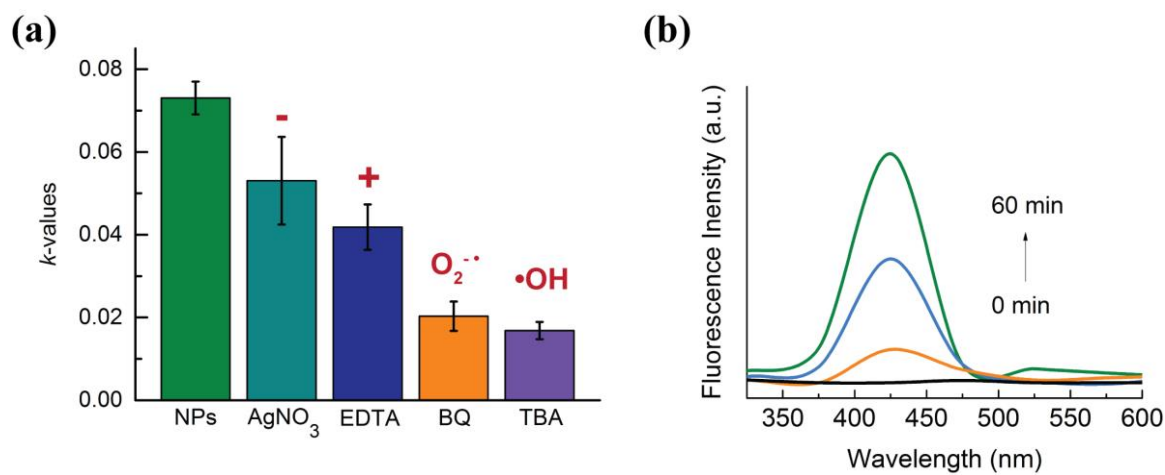

**Figure 5**. Magnetoelectric effect induced catalytic degradation mechanism of RhB using core-shell CFO-BFO NPs under 15 mT and 1.1 kHz magnetic fields. (a) Trapping experiments demonstrating the effect of the four reactive species on degradation efficiency of RhB, where, -, +, $O_2^{\cdot-}$ and $OH^{\cdot}$ refer to negative charge carriers, positive charge carriers, superoxide and hydroxyl radicals, respectively. (b) Reaction of terephthalic acid with $OH^{\cdot}$ radicals to produce increasing amounts of fluorescent 2-hydroxyterephthalic acid with peak intensity at 425 nm.

**Table 1.** List of all the scavengers used in the trapping experiments and the reactive species they quench.

| Scavenger | Reactive species quenched |
| --- | --- |
| AgNO$_3$ | e$^-$ |
| Ethylene diamine tetraacetic acid (EDTA) | h$^+$ |
| Tert-butyl alcohol (TBA) | OH$^\bullet$ |
| Benzoquinone (BQ) | O$_2^{\bullet-}$ |

**Organic compounds are degradaded by the magnetoelectric (ME) nature of cobalt ferrite-bismuth ferrite (CFO-BFO) core-shell nanoparticles.** The combination of magnetostrictive CFO with the multiferroic BFO gives rise to a magnetoelectric engine that purifies water under wireless magnetic fields via advanced oxidation processes, without involvement of any sacrificial molecules or co-catalysts.

**magnetoelectric, multiferroic, bismuth ferrite, catalysis, organics degradation**

F. Mushtaq*, X.-Z. Chen*, H. Torlakcik, C. Steuer, M. Hoop, E. C. Siringil, Xavi Marti, G. Limburg, P. Stipp, B. J. Nelson, S. Pané*

**Magnetoelectrically driven catalytic degradation of organics**

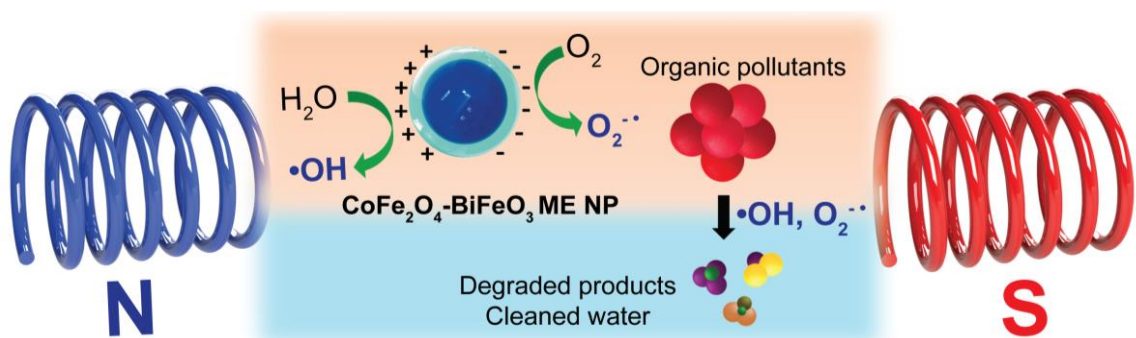

# Supporting Information

**Magnetoelectrically driven catalytic degradation of organics**

*Fajer Mushtaq\*, Xiangzhong Chen\*, Harun Torlakcik, Christian Steuer, Marcus Hoop, Erdem Can Siringil, Xavi Marti, Gregory Limburg, Patrick Stipp, Bradley J. Nelson and Salvador Pané\**

**Experimental Section**

**Fabrication of core-shell $CoFe_2O_4$-$BiFeO_3$ nanostructures**

$CoFe_2O_4$ (CFO) nanoparticles (NPs) were fabricated by a hydrothermal synthesis approach. For the fabrication of CFO NPs, 0.14 M hexadecyltrimethylammonium bromide (CTAB), 0.092 M $FeCl_3 \cdot 6H_2O$ and 0.046 M $CoCl_2$ were dissolved in DI water under continuous mechanical stirring. Next, a 6 M NaOH solution was added to the above solution under vigorous mechanical stirring followed by ultrasound. Finally, the above solution was transferred to a sealed, Teflon-lined steel autoclave and heated at 180 ˚C for 24 h. The obtained black powder was washed with DI water and ethanol and dried overnight at 80 ˚C. Next, a precursor of $BiFeO_3$ (BFO) was prepared by dissolving 0.011 M $Bi(NO_3)_3 \cdot 5H_2O$ and 0.01 M $Fe(NO_3)_3 \cdot 9H_2O$ in ethylene glycol. $CoFe_2O_4$-$BiFeO_3$ core-shell nanostructures were prepared by dispersing 0.1 g of dried CFO nanoparticles into 60 mL of the BFO precursor solution and sonicated for 2 h. This solution was then dried at 80 ˚C overnight, followed by annealing the dried powder at 600 ˚C for 2 h at a heating ramp rate of 10 ˚C min$^{-1}$.

**Material characterization**

Morphology of the resulting CFO NPs was studied by transmission electron microscopy (TEM, FEI F30), and scanning transmission electron microscopy (STEM, FEI F30). Distribution of elements along the nanoparticles were studied by energy-dispersive X-ray (EDX) mapping using HAADF STEM (FEI Talos F200X). The crystallographic structure of the nanostructures was analyzed by X-ray diffraction (XRD) on a Bruker AXS D8 Advance

X-ray diffractometer, equipped with a Cu target with a wavelength of 1.542 Å. Local crystallographic structure was studied by selected area electron diffraction (SAED). Piezoresponse force microscopy (PFM) investigations were performed on a commercial atomic force microscope (NT-MDT Ntegra Prima). PtIr-coated Si probes, i.e. FMG01/Pt (spring constant k ~ 3 N m$^{-1}$), were used, and the imaging contact force set-points were carefully controlled. To acquire local piezoresponse loops, ac signals ($V_{AC}$ = 0.5 V) were superimposed on triangular staircase wave with DC switching from -10 V to 10 V. To study change in piezoelectric response of the sample under magnetic field, an in-plane magnetic field of 1000 Oe was applied to the sample.

**Multiphysics simulation of ME CFO-BFO NPs**

The simulations were implemented in the commercially available software COMSOL Multiphysics based on similar examples from literature.[1,2] The physics of our COMSOL model used to describe the ME effect was divided into magnetic fields, solid mechanics and electrostatics. Influences from the surrounding medium on the induced electrical surface potential are neglected and the relative permittivity and permeability is assumed to be 1. For these simulations presented in Figure 3 c-d, octahedral CFO NP with a diameter of 30 nm and a BFO shell with a thickness of 5 nm were selected as input parameters. An epitaxially grown BFO shell on the CFO core´s [111] plane was considered and implemented accordingly in the model.[3,4] The magnetic field strength was fixed at 15 mT and applied on the boundaries of the medium along the global z-axis. Since the NPs are free to move in the surrounding medium, it was assumed that under magnetic field they align with the excitation field and hence, magnetostriction along the easy axis was considered. COMSOL was used to compute the magnetization gradients within the material by using the applied magnetic field and its corresponding magnetization values from the measured VSM hysteresis curve for CFO NPs (Figure S6). The internal strain generated in a CFO NP under magnetic fields was governed by the following equation,

$$\lambda_z = 1.5\lambda_s \frac{M_z}{M_{sat}}$$

where, $\lambda_z$ is the strain along the z-axis, $\lambda_s$ the magnetostriction parameter ($\lambda_{s,CFO}$= -273 ppm, $\lambda_{s,BFO}$=-0.002 ppm), $M_z$ the magnetization along the z-axis and $M_{sat}$ the saturation magnetization of the CFO core.[5-7] The strain transfer from CFO core to the BFO shell was assumed to be ideal.[1] This strain in the BFO shell is converted into electric polarization on the surface of BFO. BFOs' piezo-electric coupling and compliance matrix with R3c symmetry are given by

$$d = \begin{bmatrix} 13.5 & 0 & 0 & 0 & 9 & 0 \\ 0 & 13.5 & 0 & 9 & 0 & 0 \\ 3 & 3 & 50 & 0 & 0 & 0 \end{bmatrix} \cdot 10^{-12} [C/N]$$

$$s_E = \begin{bmatrix} 0.0182167 & -0.0006753 & -0.0179691 & 0.0011707 & 0 & 0 \\ \cdot & 0.0182167 & -0.0179691 & -0.0011707 & 0 & 0 \\ \cdot & \cdot & 0.0492076 & 0 & 0 & 0 \\ \cdot & \cdot & \cdot & 0.0192123 & 0 & 0 \\ \cdot & \cdot & \cdot & \cdot & 0.0192123 & 2 \cdot 0.0011707 \\ \cdot & \cdot & \cdot & \cdot & \cdot & 0.0377839 \end{bmatrix} \cdot 10^{-9} [1/Pa]$$

and were derived from literature.[2,8-10]

The remaining elastic and electric properties of the CFO core and BFO shell were also determined from literature.[2,11-15] The mechanical boundary condition was set in the middle plane of the CFO core through fixing the vertices. The electrical ground was applied on the boundaries of the medium. For the study of the induced electrical surface potential as a function of BFO shell thickness, shell thickness values were selected from 2.5 nm to 20 nm.

**Magnetoelectric effect induced RhB degradation measurement**

Experiments were performed to study the degradation of RhB dye in the presence of our CFO-BFO NPs using AC magnetic fields. An RhB concentration of 2 mg L$^{-1}$ was chosen to perform degradation experiments. 20 mg of CFO-BFO NPs were dispersed in 20 mL of RhB solution and placed inside the custom-built magnetic set-up and subjected to various magnetic fields and frequencies under constant agitation, after the adsorption-desorption equilibrium was reached. A UV-Vis spectrophotometer (Tecan Infinite 200 Pro) was used to obtain the fluorescent spectra of RhB over time by taking aliquots of irradiated RhB solution every 10 minutes for 50 minutes.

**Trapping experiments**

To investigate the degradation pathway behind magnetoelectric effect-induced catalysis, we performed trapping experiments by using different scavengers. AgNO$_3$ (2 mM), ethylene diamine tetraacetic acid (EDTA, 2 mM), tert-butyl alcohol (TBA, 2 mM) and benzoquinone (BQ, 0.5 mM) solutions were prepared in a 2 mg L$^{-1}$ RhB solution. For the catalysis experiments, 2 mg of CFO-BFO NPs were dispersed in 2 mL of RhB solution and placed inside the custom-built magnetic set-up. To probe the formation of OH$^{\bullet}$ radicals, 0.5 mM terephthalic acid solution was prepared and subjected to AC magnetic fields with CFO-BFO NPs, after which the solution's intensity was monitored at 425 nm every 30 mins.

**Magnetoelectric effect induced micro-pollutant degradation measurement**

Experiments were performed to study the degradation of five common pharmaceuticals in the presence of our CFO-BFO NPs using AC magnetic fields. The pharmaceuticals chosen for this study were Carbamazepine, Diclofenac, Gabapentin, Oxazepam and Fluconazole, each having a concentration of 50 µg L$^{-1}$. 50 mg of CFO-BFO NPs were dispersed in 55 mL of above solution and placed inside the custom-built magnetoelectric set-up and subjected to the optimised magnetic field conditions under constant agitation, once the adsorption-desorption equilibrium was reached. After magnetic treatment, the NPs were removed from the solution using centrifugation and magnetic separation. To compare the efficiency of the magnetic treatment, a second sample set was prepared without being subjected to any magnetic treatment. Sample preparation for evaluating concentration of pharmaceuticals was done as described by Dasenaki et al. .[16] Briefly, 50 mL of samples were used. Samples were stored at 4 °C and were analyzed within 5 days. All samples showed a pH value between 1.9 and 2.1 and hence, samples were not acidified as reported before.[16] Before loading, Strata-XL cartridge were first conditioned with 6 mL of methanol and afterwards with 6 mL of ultrapure water. Conditioning process was done under gravity. Samples were loaded on the cartridges, also under gravity. After loading, cartridges were washed with 6 mL of pure water and subsequently dried under reduced pressure for 30 min. Target analytes were eluted with 6 mL of methanol. Solvent was evaporated under a gentle steam of nitrogen at 40 °C. Dried residues were dissolved in 100 µL of 0.1% formic acid in water. Five replicates of each sample were analyzed. Analysis was performed using a LTQ-XL linear ion trap (Thermo Scientific, San Jose (CA), United States ) mass spectrometer coupled to a Waters Acquity™ UPLC system (Milford (MA), United States). Gradient elution was done on a Waters BEH C18 column (2.1 x 50 mm, 1.7 µm). The mobile phase consisted of 0.1% formic acid in water (eluent A) and formic acid (0.1%) in acetonitrile (eluent B). The flow rate was set to 0.5 mL/min and the injection volume was 10 µL. Dwell volume of the UPLC system was 0.7 mL. The final LC gradient was as follows: 0-2 min 2% B, 2-12 min to 90% B, 12-15 90% B, 15-15.5 min to 2% B,15.5 -20 min 2% B. The column oven and the autosampler was set to 30 °C and 10 °C, respectively. MS settings were used as reported by Wissenbach et al.[17] and are as follows. Analysis of carbamazepine, fluconazole, gabapentine and oxazepam was performed in the positive ionization mode. For diclofenac, negative ionization mode under same solvent conditions was used. The LTQ-XL was equipped with a heated ESI II source set to 150 °C. Sheath gas 40 arbitrary units (AU), auxiliary gas 20 AU: source voltage 3.00 kV; ion transfer capillary 300 °C. capillary voltage 31 V; tube lens voltage 80 V. Automatic gain control was set to 15000 ions for full scan and 5000 for $MS^n$. Collision induced dissociation (CID)- $MS^n$ experiments were performed on precursor ions selected for $MS^1$. Using information dependent acquisition, $MS^1$ was performed was performed in full scan mode (*m/z* 100-1500). $MS^2$ and $MS^3$ were performed in the IDA mode: four IDA $MS^2$ experiments were performed on the four most intensive signal from $MS^1$ and additionally eight $MS^3$ scan filters were chosen to record the most and second most ions from $MS^2$. Removal efficiency was evaluated by comparing peak areas of the respective drugs before and after treatment.

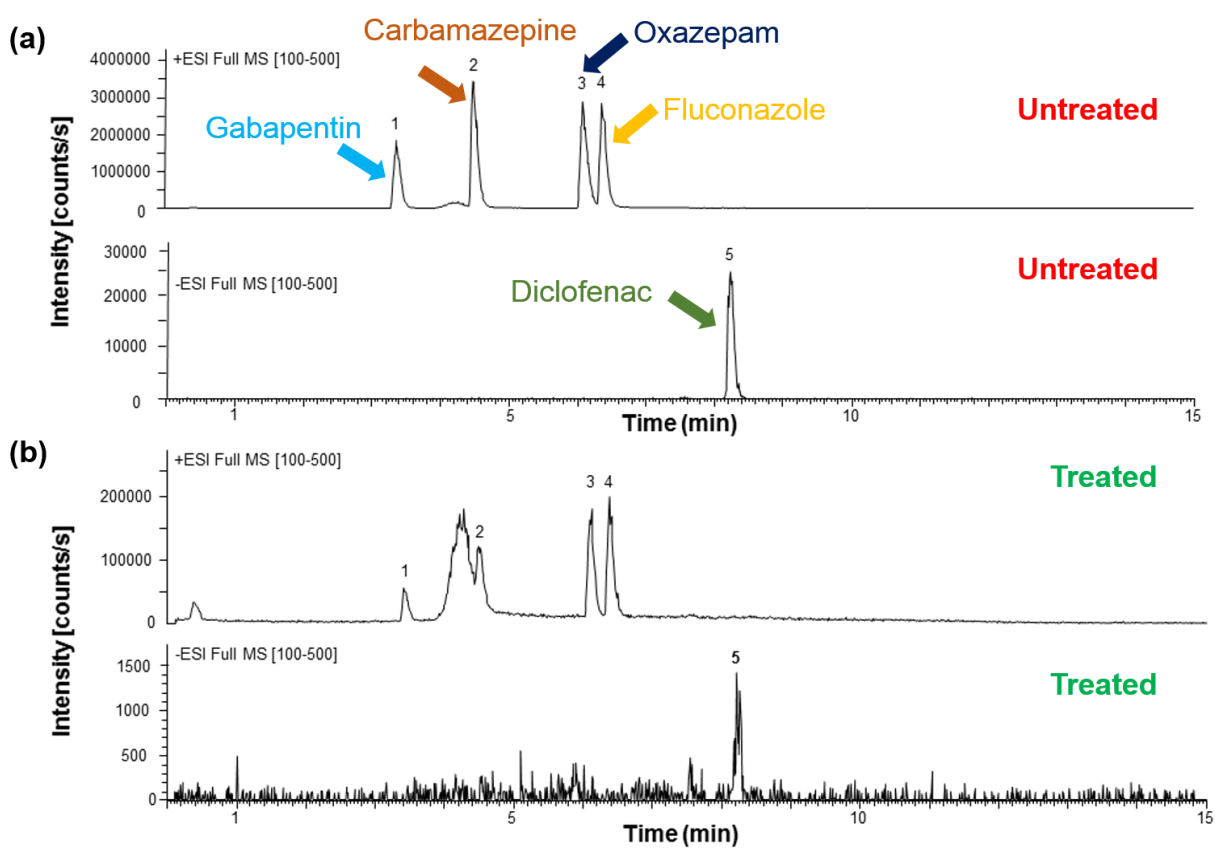

Figure S1. MS spectra obtained for the five pharmaceuticals (a) before and (b) after treatment with ME CFO-BFO NPs under magnetic fields, showing drastically reduced concentrations after treatment.

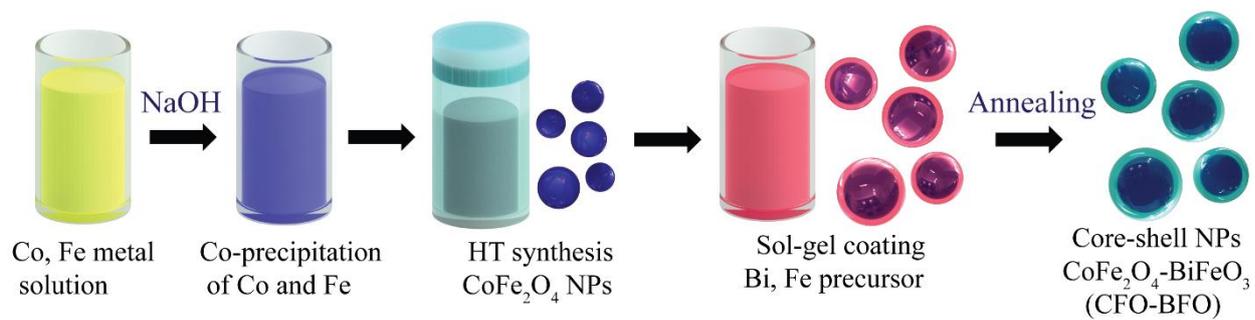

Figure S2. Fabrication scheme of CFO-BFO nanoparticles using hydrothermal and sol-gel synthesis.

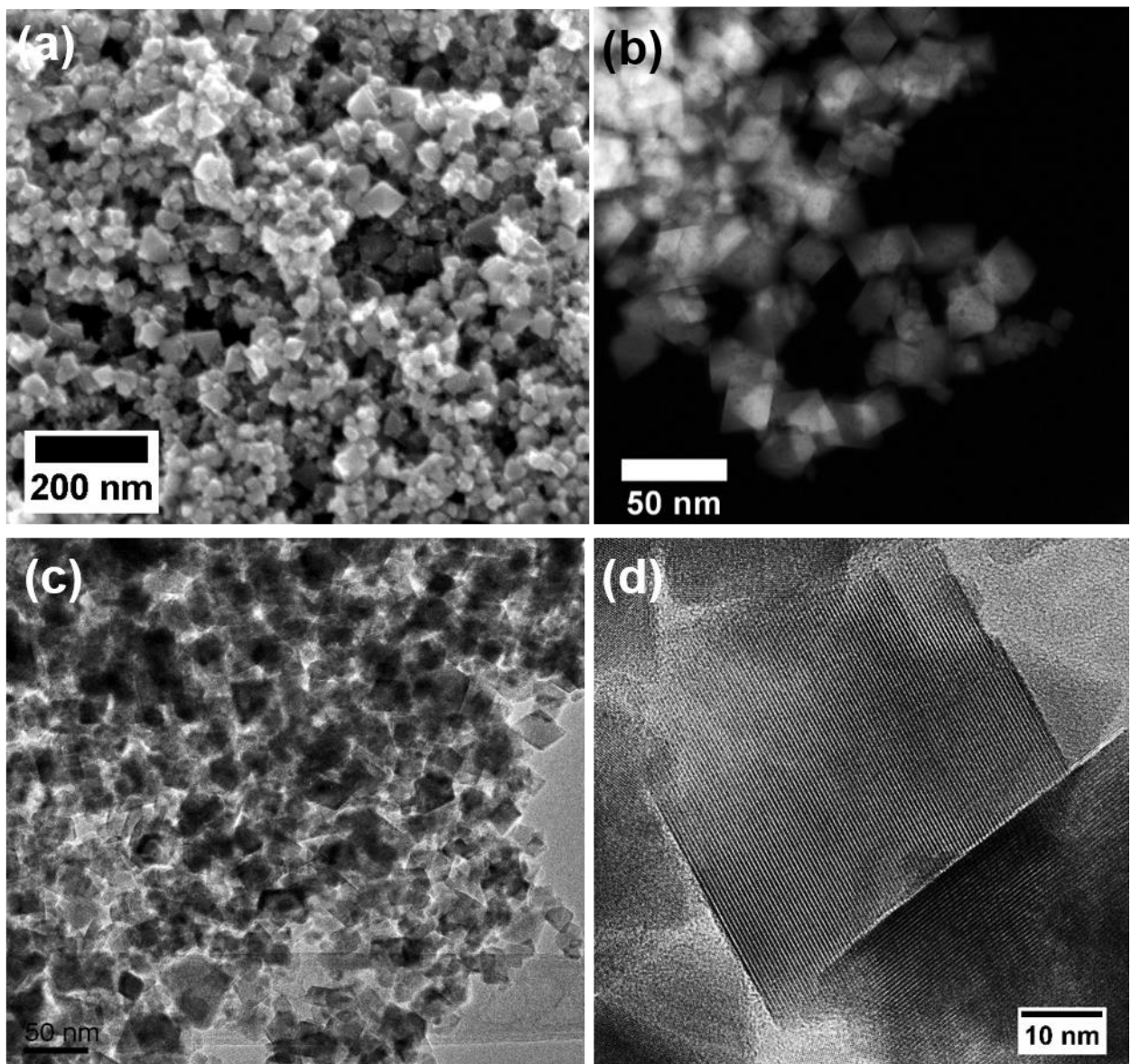

Figure S3. Characterisation of CFO NPs uisng SEM and TEM. (a) SEM image showing many octahedron CFO NPs. (b) STEM and (c) TEM image of octahedral CFO NPs. (d) HRTEM image showing a is single crystalline CFO NP.

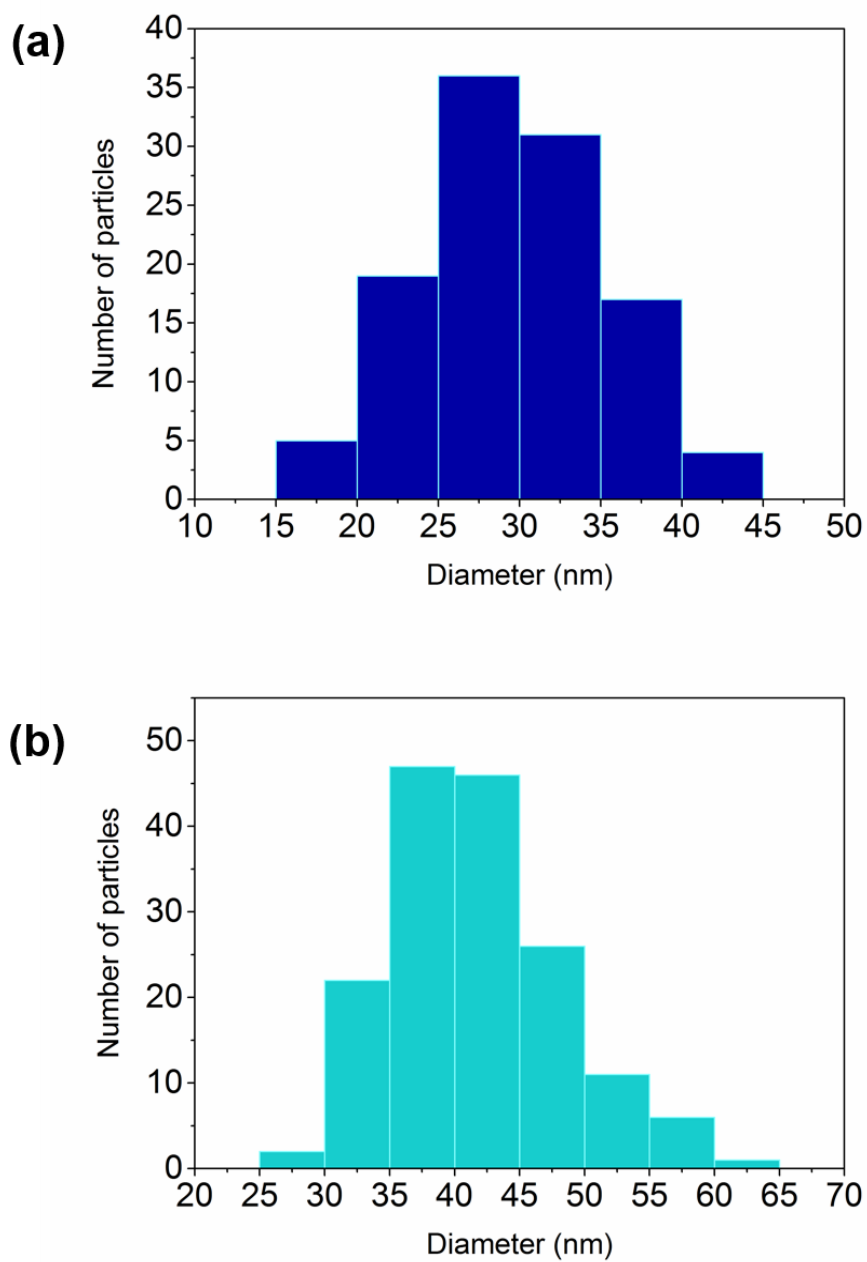

Figure S4. Size distribution of (a) CFO NPs with an average size of 30 ± 6 nm and (b) CFO-BFO NPs with an average size of 42 ± 6 nm.

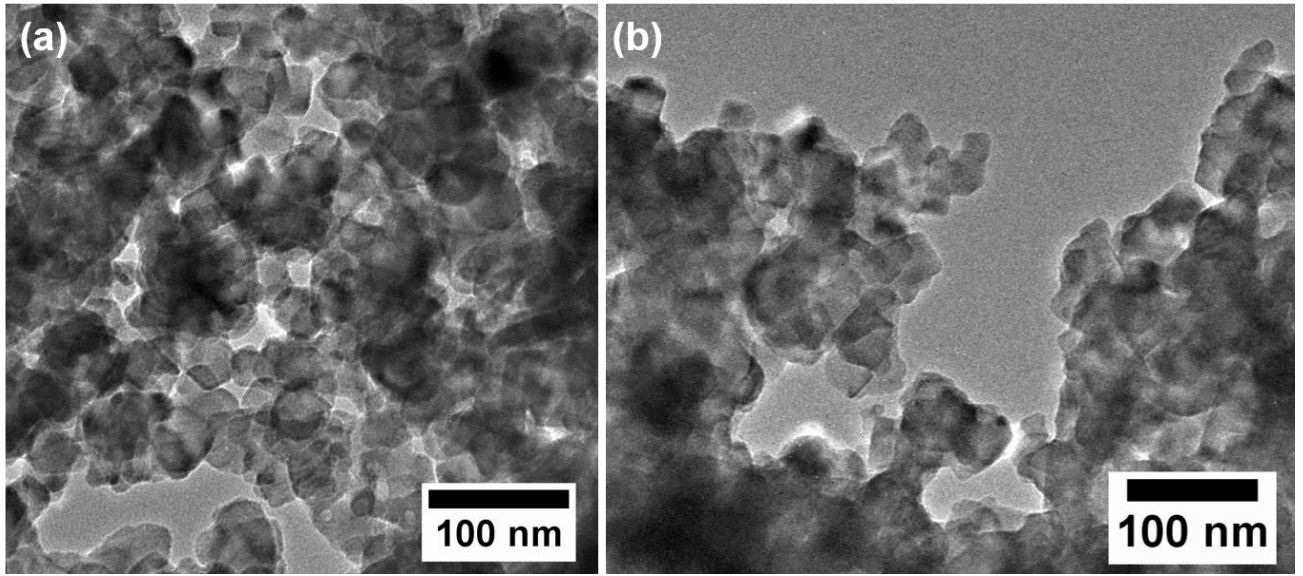

Figure S5. TEM images showing core-shell CFO-BFO NPs after sol-gel coating and annelaing.

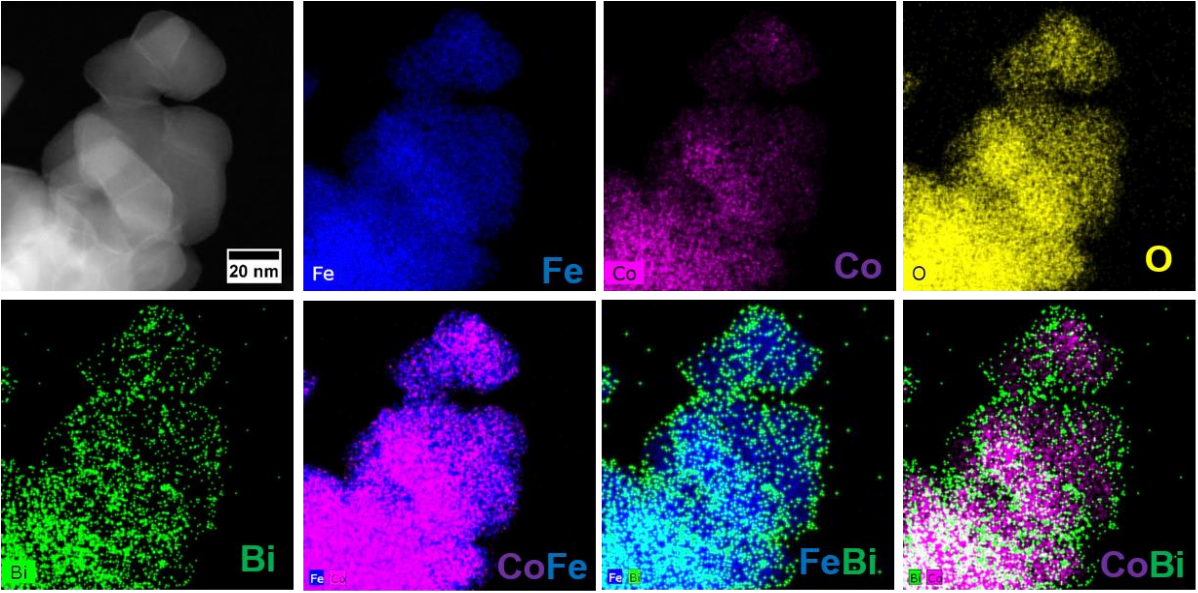

Figure S6. EDX maps obtained for HRTEM image showing CFO-BFO core-shell NPs, where the distribution of elements clearly highlights the presence of a BFO shell formed around a CFO core.

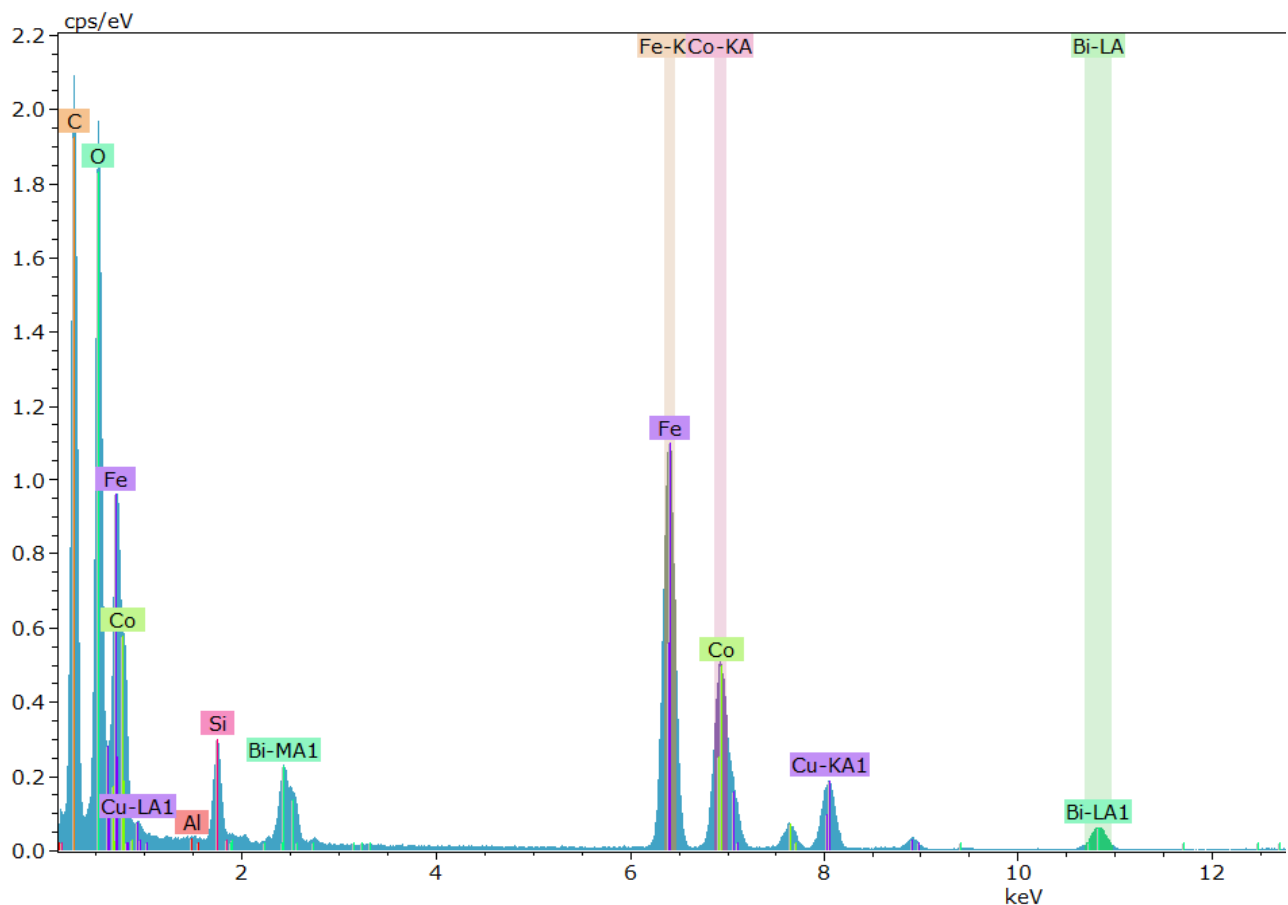

Figure S7. EDX spectra obtained during TEM analysis performed on core-shell CFO-BFO NPs showing the presence of O, Fe, Co and Bi elements. (peaks from Cu and Si are origination form the TEM sample holder).

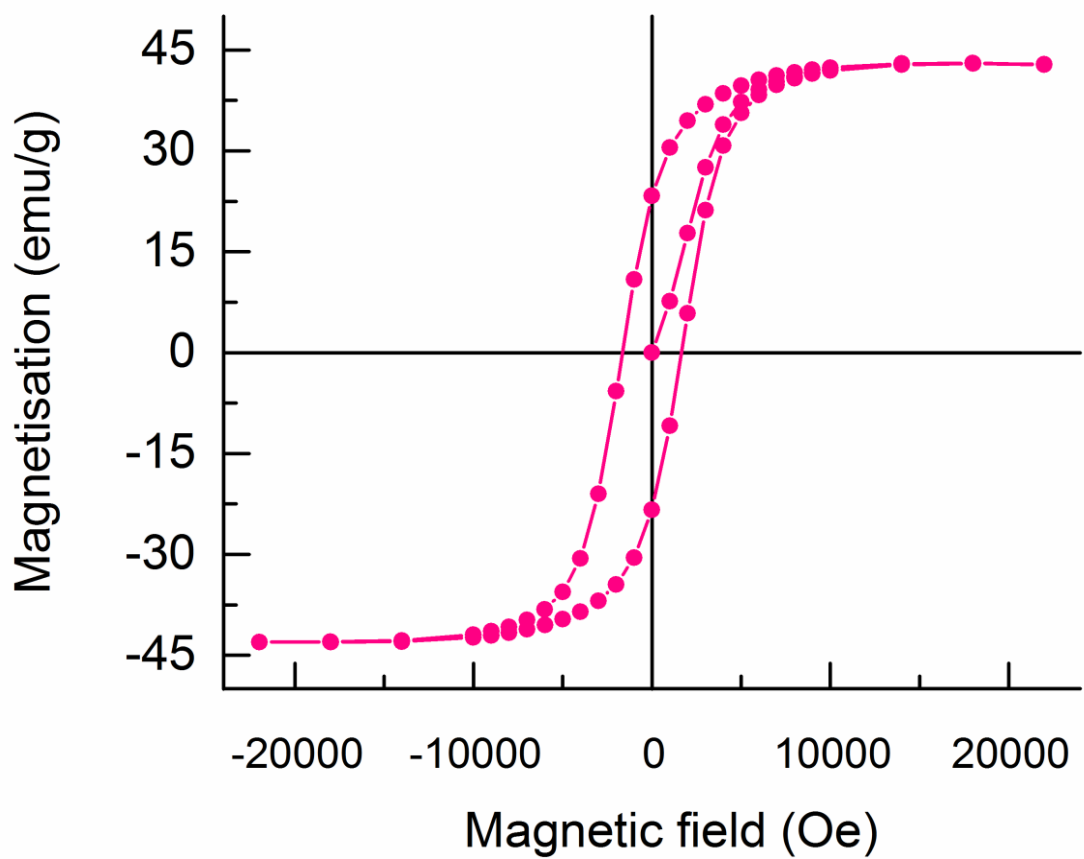

Figure S6. Magnetic characterisation of as-fabricated CFO NPs (without using magnetic separation) to investigate their initial magnetisation and magnetic saturation $M_s$ values.

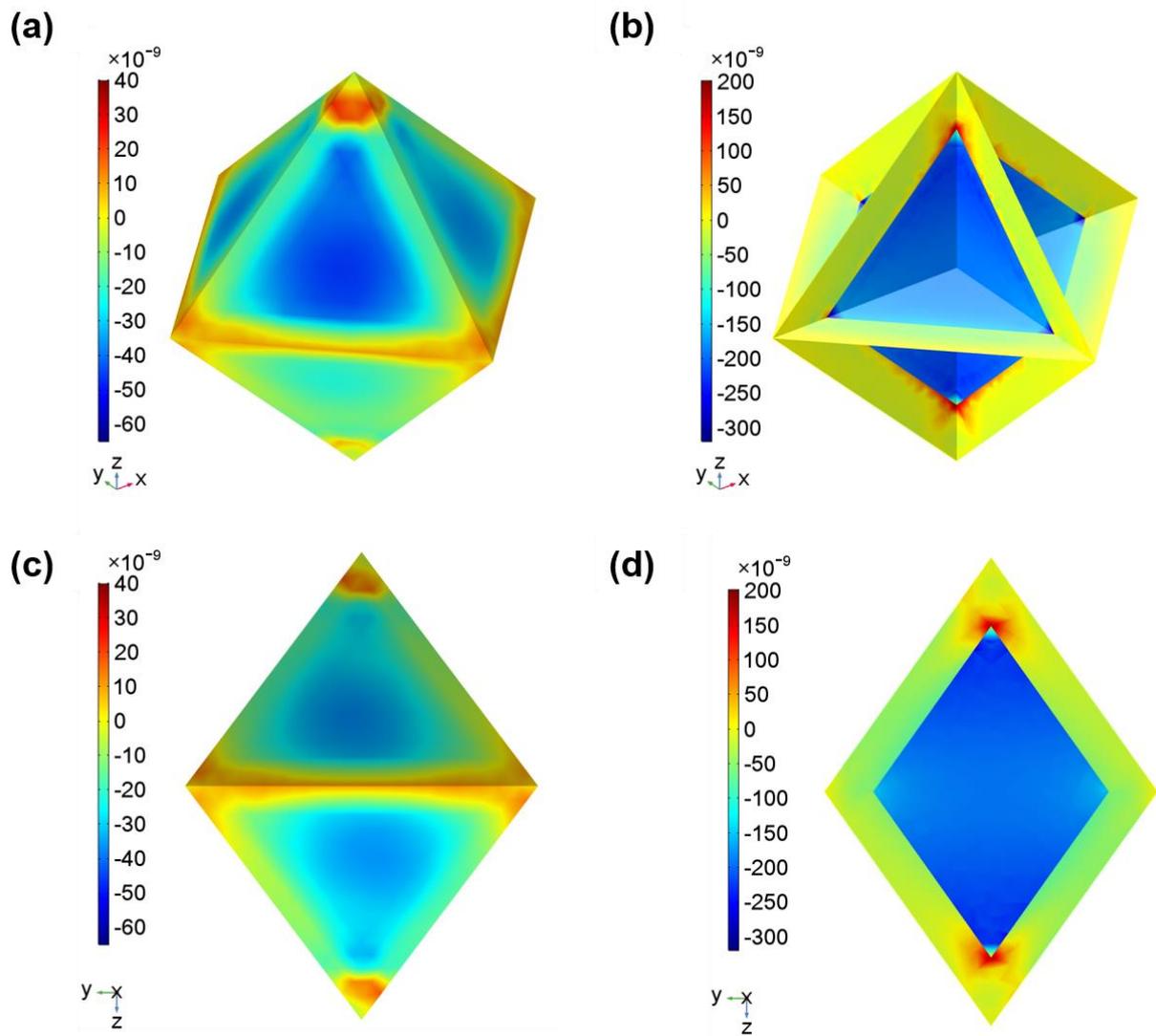

Figure S7. COMSOL multiphysics simulations on CFO-BFO NPs for a BFO shell thickness of 5 nm and an applied magnetic field of 15 mT. (a,c) volumetric strain distribution on the surface of CFO-BFO NP and (b.d) the corresponding strain distribution in cross-section.

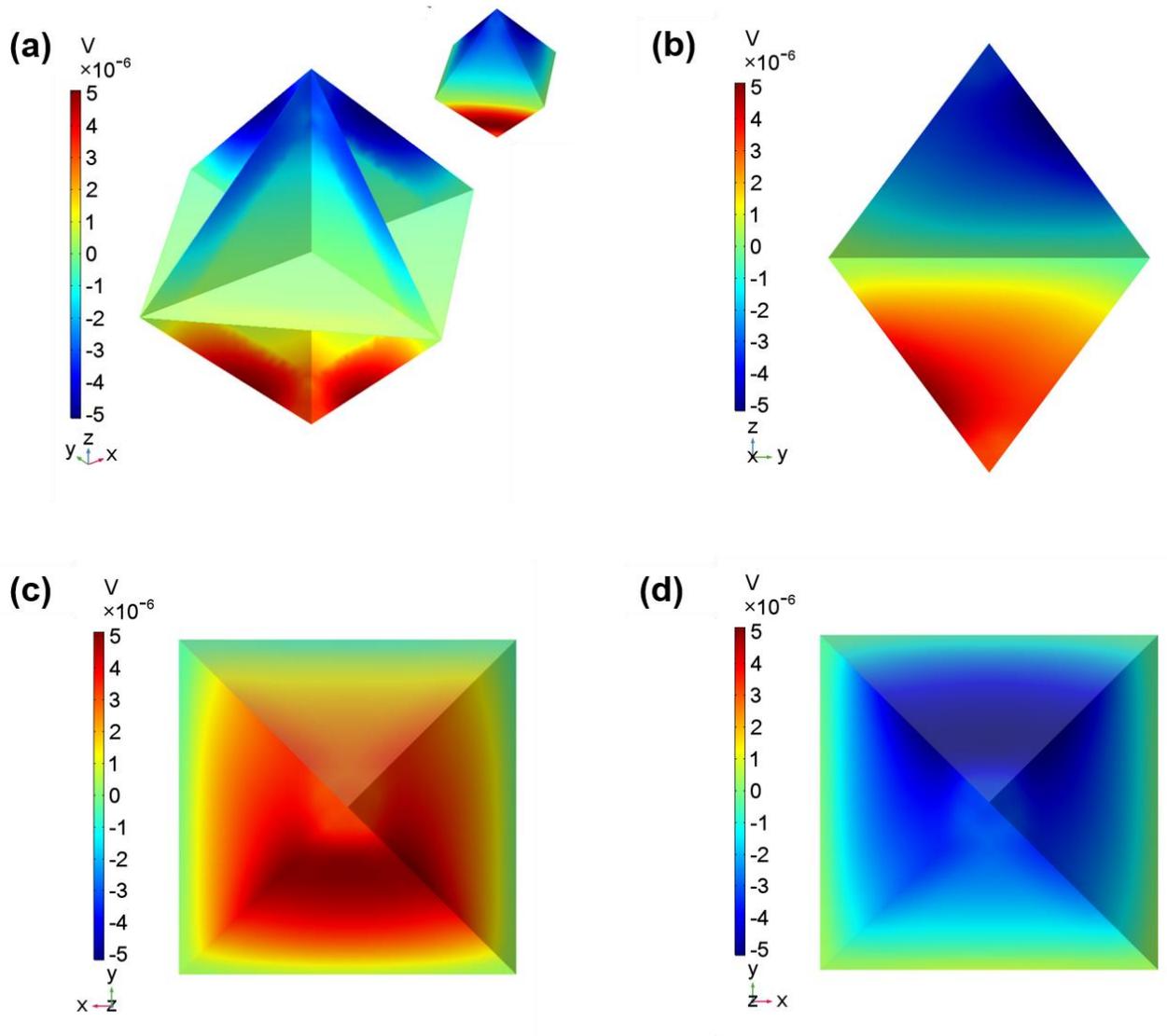

Figure S8. COMSOL multiphysics simulations on CFO-BFO NPs for a BFO shell thickness of 5 nm and an applied magnetic field of 15 mT. Potential generated (a) in the cross-section of CFO-BFO NP, (b-d) on the surface of BFO shell viewed from different angles showing a gradient potential distribution with opposite polarities on the extreme sides.

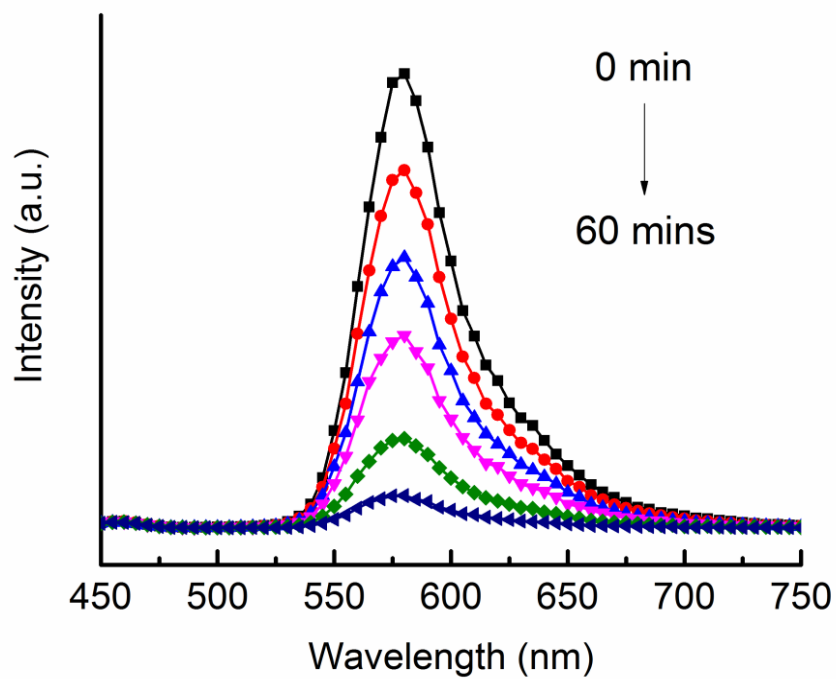

Figure S9. Fluorescent spectra of RhB collected every 10 minutes and measured at a peak of 554 nm during magnetoelectrically induced catalytic degradation of RhB. From this plot it can be seen that with increasing reaction time, RhB's concentration reduces dramatically.

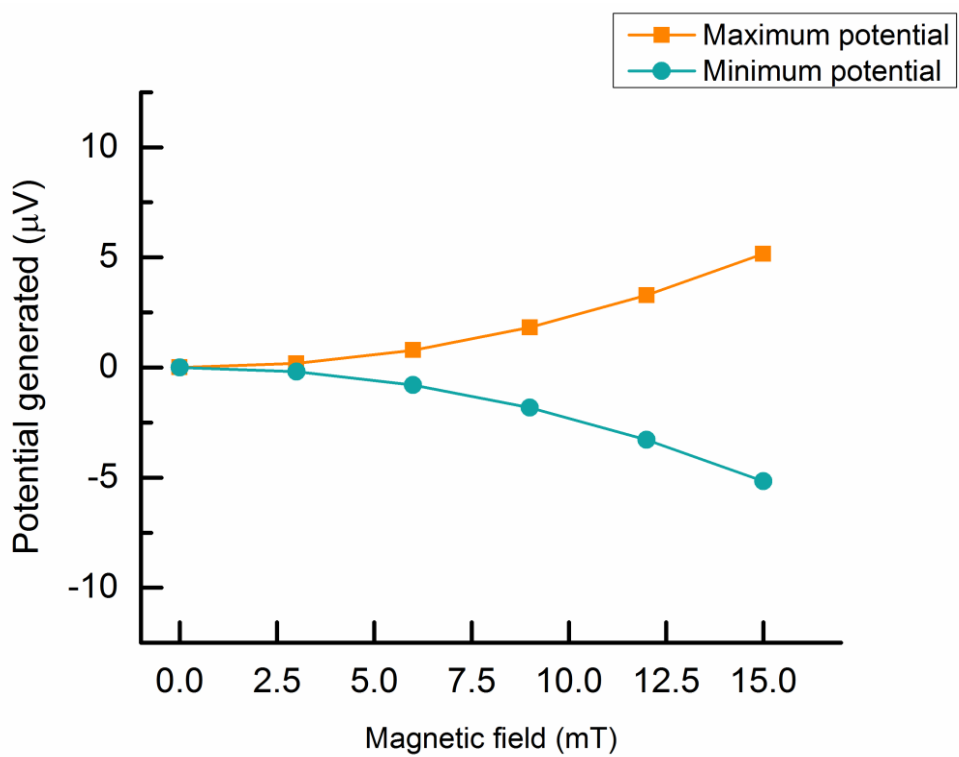

Figure S10. COMSOL simulations performed on CFO-BFO NPs to study the potential generated on BFO shell under increasing magnetic field strengths. This plot clearly shows that increasing the magnetic field leads to an increase in the induced surface potential on BFO shell, a trend that supports the RhB degradation rates presented in Figure 4b.

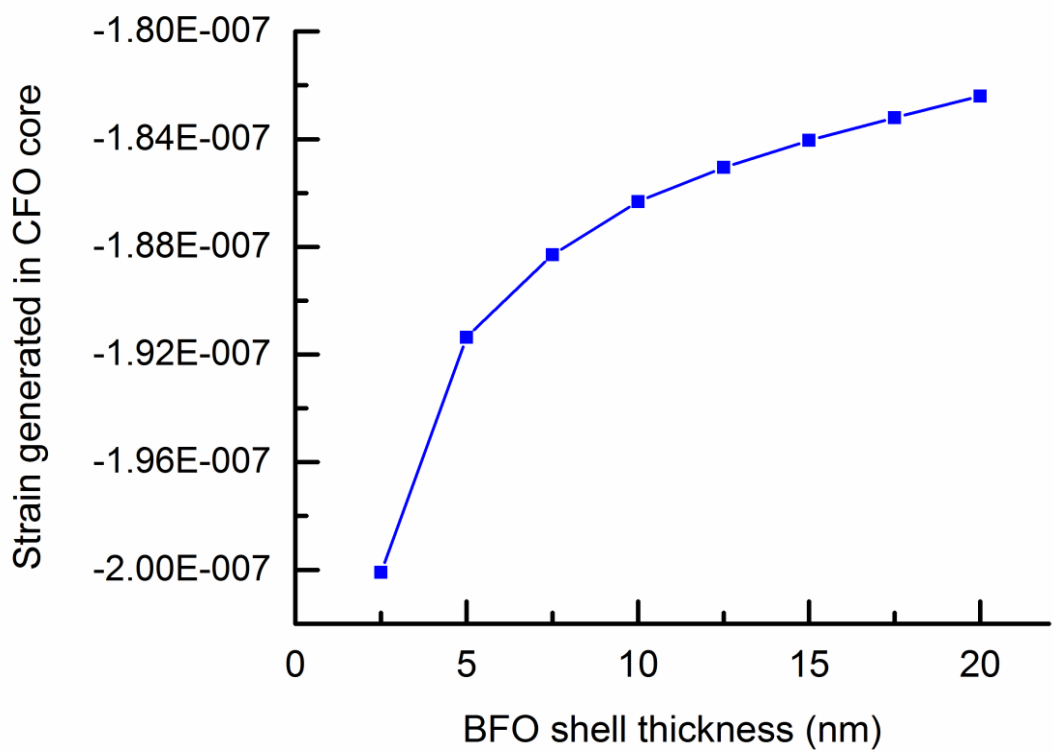

Figure S11. COMSOL simulations performed on CFO-BFO NPs with varying shell thickness show that increasing shell thickness of BFO led to a lower absolute strain generation in the CFO core, presumably due to the clamping effects.

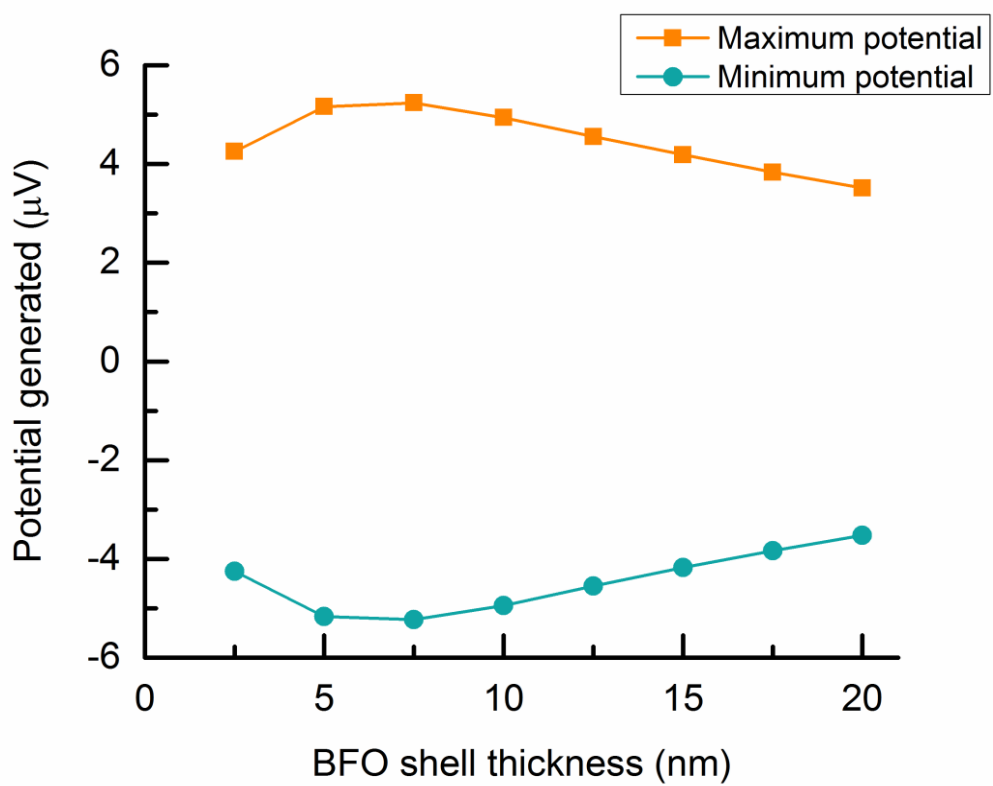

Figure S12. COMSOL simulations performed on CFO-BFO NPs with varying shell thickness and its influence on the potential generated on the BFO shell. An optimal BFO shell thickness was identified at 7.5 nm, which is close to the shell thickness of our fabricated core-shell structures, where the maximum electrical potential of 5.24 μV could be generated.

**References**

(1) Betal, S.; Shrestha, B.; Dutta, M.; Cotica, L. F.; Khachatryan, E.; Nash, K.; Tang, L.; Bhalla, A. S.; Guo, R. *Scientific Reports* **2016**, *6*, 32019.

(2) Aimon, N. M.; Liao, J.; Ross, C. A. *Applied Physics Letters* **2012**, *101*.

(3) Madigou, V.; Souza, C. P. D.; Madigou, V.; Souza, C. P. D.; Con, C. L. **2014**.

(4) Apc International, L. *Piezoelectric ceramics: principles and applications*; APC International, 2002.

(5) Chikazumi, S. *Physics of ferromagnetism*; 2nd ed., r ed.; Oxford : Oxford University Press, 1999; Vol. 94.

(6) Khaja Mohaideen, K.; Joy, P. A. *Journal of Magnetism and Magnetic Materials* **2014**, *371*, 121.

(7) Kumar, A.; Scott, J. F.; Martínez, R.; Srinivasan, G.; Katiyar, R. S. *Physica Status Solidi (A) Applications and Materials Science* **2012**, *209*, 1207.

(8) Graf, M.; Sepliarsky, M.; Machado, R.; Stachiotti, M. G. *Solid State Communications* **2015**, *218*, 10.

(9) Wang, Y. L.; Wu, Z. H.; Deng, Z. C.; Chu, L. Z.; Liu, B. T.; Liang, W. H.; Fu, G. S. *Ferroelectrics* **2009**, *386*, 133.

(10) Nye, J. F. *Physical properties of crystals : their representation by tensors and matrices*; [Reprinted ed.; Oxford : Clarendon Press, 2009.

(11) Catalan, G.; Scott, J. F. *Advanced Materials* **2009**, *21*, 2463.

(12) Bueno-Baques, D.; Corral-Flores, V.; Morales-Carrillo, N. A.; Torres, A.; Camacho-Montes, H.; Ziolo, R. F. *Mater. Res. Soc. Symp. Proc* **2017**.

(13) Gujar, T. P.; Shinde, V. R.; Kulkarni, S. S.; Pathan, H. M.; Lokhande, C. D. *Applied Surface Science* **2006**, *252*, 3585.

(14) Vasundhara, K.; Achary, S. N.; Deshpande, S. K.; Babu, P. D.; Meena, S. S.; Tyagi, A. K. *Journal of Applied Physics* **2013**, *113*.

(15) Gujar, T. P.; Shinde, V. R.; Lokhande, C. D. *Materials Chemistry and Physics* **2007**, *103*, 142.

(16) Dasenaki, M. E.; Thomaidis, N. S. *Anal. Bioanal. Chem.* **2015**, *407*, 4229.

(17) Wissenbach, D. K.; Meyer, M. R.; Weber, A. A.; Remane, D.; Ewald, A. H.; Peters, F. T.; Maurer, H. H. *J. Mass Spectrom.* **2012**, *47*, 66.